\documentclass[aps,prd, notitlepage, onecolumn,superscriptaddress,floatfix,letterpaper,nofootinbib, longbibliography]{revtex4-1}

\usepackage{amsmath, amsthm, amssymb,slashed}

\usepackage[usenames, dvipsnames]{color}
\usepackage[svgnames]{xcolor}
\usepackage[colorlinks,citecolor=RoyalBlue, urlcolor=RoyalBlue, linkcolor=RoyalBlue ]{hyperref} 

\usepackage[normalem]{ulem}

\usepackage{booktabs}

\newcommand{\ccblue}[1]{\textcolor{black}{#1}}

\definecolor{mygray}{gray}{0.6}

\usepackage{upgreek}
\usepackage{bbm}

\newenvironment{myfont}[2][]{\csname#2\endcsname[#1]}{}

\usepackage{slashed}
\usepackage[makeroom]{cancel}
\usepackage[normalem]{ulem}
\usepackage{soul}
\newcommand{\stkout}[1]{\ifmmode\text{\sout{\ensuremath{#1}}}\else\sout{#1}\fi}

\usepackage{sseq}
\usepackage[all,cmtip]{xy}
\usepackage{tikz-cd}
\usepackage{tikz}
\usetikzlibrary{matrix}
\usetikzlibrary{decorations.markings}
\usetikzlibrary{tikzmark,decorations.pathreplacing,positioning}
\usepackage{amsfonts}

\newcommand{\bea}{\begin{eqnarray}}
\newcommand{\eea}{\end{eqnarray}}
\def\be{\begin{equation}}
\def\ee{\end{equation}}

\newcommand{\e}{\hspace{1pt}\mathrm{e}}

\newcommand{\ii}{\hspace{1pt}\mathrm{i}\hspace{1pt}}

\definecolor{red}{rgb}{1,0,0}
\definecolor{blue}{rgb}{0,0,1}
\definecolor{dblue}{rgb}{0,0,0.4}
\definecolor{green}{rgb}{0,1,0}
\definecolor{black}{rgb}{0,0,0}
\definecolor{white}{rgb}{1,1,1}

\definecolor{brn}{rgb}{.8,.4,.0}
\definecolor{redo}{rgb}{1,.5,.0}
\definecolor{ddgrn}{rgb}{0,0.4,0}
\definecolor{dgrn}{rgb}{0,0.55,0}
\definecolor{dbl}{rgb}{0,0,0.5}

\usepackage[bbgreekl]{mathbbol}
\usepackage{amscd}

\newcommand{\Z}{\mathbb{Z}}

\newcommand{\R}{\mathbb{R}}

\newcommand{\dd}{\mathrm{d}}

\newcommand{\Refe}[1]{Ref.~[\onlinecite{#1}]}
 
\newcommand{\eq}[1]{eq.~(\ref{#1})} 
\newcommand{\eqq}[1]{(\ref{#1})}

\newcommand{\Tr}{{\rm Tr}}

\newcommand{\prt}{\partial}

\newcommand{\bpm}{\begin{pmatrix}}
\newcommand{\epm}{\end{pmatrix}}
\newcommand{\bmm}{\begin{matrix}}
\newcommand{\emm}{\end{matrix}}

\newcommand{\cP}{ {\cal P} }

\def\Z{{\mathbb{Z}}}

\def\R{{\mathbb{R}}}

\def\Tr{{\mathrm{Tr}}}

\def \H{\operatorname{H}}

\def \Z{\mathbb{Z}}

\newcommand {\emptycomment}[1]{}

\def\TP{\mathrm{TP}}

\def\B{\mathrm{B}}

\newcommand{\SO}{{\rm SO}}
\newcommand{\Spin}{{\rm Spin}}
\newcommand{\U}{{\rm U}}
\newcommand{\SU}{{\rm SU}}

\newcommand{\Pin}{{\rm Pin}}

\def\bZ{{\mathbf{Z}}}
\usepackage{centernot}

\newcommand{\Sec}[1]{Sec.~\ref{#1}} 

\usepackage{enumitem} 
\usepackage{mathtools,amssymb,varwidth}

\usepackage{datetime}

\newcommand{\SM}{{\rm SM}} 
\newcommand{\PD}{{\rm PD}}

\newcommand{\rmod}{\;{\rm mod}\;} 

\newcommand{\rCS}{{\rm CS}} 
\newcommand{\rA}{{\rm A}}

\begin{document}


\title{Cobordism and Deformation Class of the Standard Model 
}

\author{Juven Wang}
\affiliation{Center of Mathematical Sciences and Applications, Harvard University, MA 02138, USA}

\author{Zheyan Wan}
\affiliation{Yau Mathematical Sciences Center, Tsinghua University, Beijing 100084, China}

\author{Yi-Zhuang You}
\affiliation{Department of Physics, University of California, San Diego, CA 92093, USA}

\begin{abstract} 

't Hooft anomalies of quantum field theories (QFTs) with an invertible global symmetry $G$ (including spacetime and internal symmetries)
in a $d$d spacetime are known to be classified by a $d+1$d cobordism group TP$_{d+1}(G)$, 
whose 
group generator is a $d+1$d cobordism invariant written as an invertible topological field theory (iTFT) with a partition function $\bZ_{d+1}$. 
It has recently been proposed that the deformation class of QFTs is specified by its symmetry $G$ and an iTFT $\bZ_{d+1}$. 
%
Seemingly different QFTs of the same deformation class 
can be deformed to each other via quantum phase transitions.
In this work, we ask which cobordism class and deformation class control the 4d 
Standard Model (SM) of ungauged or gauged $(\SU(3) \times \SU(2)  \times \U(1))/\Z_q$ group for $q=1,2,3,6$ with
a continuous or discrete baryon minus lepton $\mathbf{B}-  \mathbf{L}$ like symmetry.
We show that the answer contains some combination of 5d iTFTs: 
two $\mathbb{Z}$ classes associated with $({ \mathbf{B}-  \mathbf{L}})^3$ and $({ \mathbf{B}-  \mathbf{L}})$-(gravity)$^2$ 4d perturbative local anomalies,
a $\mathbb{Z}_{16}$ class  
Atiyah-Patodi-Singer $\eta$ invariant and
a $\mathbb{Z}_2$ class 
Stiefel-Whitney $w_2w_3$ invariant associated with 4d nonperturbative global anomalies, 
and additional 
$\mathbb{Z}_3 \times \mathbb{Z}_2$ 
global anomalies involving higher symmetries whose charged objects 
are Wilson electric or 't Hooft magnetic line operators.
Out of the multiple infinite $\Z$ classes of local anomalies and 24576 classes of global anomalies, 
we pin down the deformation class of the SM labeled by 
$(N_f, n_{\nu_{R}}, $ p$',q)$,
the family number, 
the total ``right-handed sterile'' neutrino number, 
the magnetic monopole datum, and 
the mod $q$ relation.
We show that 
Grand Unification
such as Georgi-Glashow $su(5)$, 
Pati-Salam $su(4) \times su(2) \times su(2)$, 
Barr's flipped $u(5)$, 
and   
the familiar or modified $so(n)$ models of Spin($n$) gauge group, e.g., with $n=10,18$
can all reside in an appropriate SM deformation class.
We show that Ultra Unification, which replaces some of sterile 
neutrinos with new exotic gapped/gapless sectors (e.g., topological or conformal field theory)
or gravitational sectors with topological origins
via cobordism constraints,
also resides in an SM deformation class.
Neighbor quantum phases near SM or their phase transitions, and neighbor gapless quantum critical regions naturally exhibit beyond SM phenomena. 

\end{abstract}


\maketitle



\section{Introduction and Summary}

Symmetries in quantum field theories (QFTs)
in the strict sense mean that the QFT is invariant under certain sets of transformations (i.e., symmetry transformations).
The symmetry invariance of QFT may be exhibited in many ways depending on how the QFT is formulated: 
(1) in the path integral formulation (say in $d$d for $d$-dimensional spacetime), 
the symmetry transformation on quantum fields or spacetime coordinates leaves the path integral $\bZ_d$ invariant,
(2) in the Hamiltonian approach, the symmetry transformation corresponds to a quantum operator $\hat U$ commuting with the Hamiltonian $\hat H$, i.e., $[\hat H, \hat U]=0$.
Importantly,  
the symmetry may only be invariant up to a global complex phase; for example,
the symmetry transformation can map the path integral $\bZ_d \mapsto \e^{\ii \Theta} \bZ_d$.
If $\e^{\ii \Theta}$ cannot be eliminated by $d$d local counter terms, 
this $\e^{\ii \Theta}$ manifests a mild violation of symmetry invariance.
Such a symmetry is an \emph{anomalous symmetry}  (i.e., symmetry with \emph{'t Hooft anomaly} \cite{tHooft1979ratanomaly}).
In contrast, any symmetry transformation without outputting any $\e^{\ii \Theta}$ is an \emph{anomaly-free symmetry}.
Note that $\e^{\ii \Theta}$ is invertible (i.e., $\e^{\ii \Theta} \cdot \e^{-\ii \Theta}=1$)
--- the \emph{anomaly inflow} \cite{1984saCallanHarvey, Witten2019bou1909.08775} states 
that the $\e^{\ii \Theta}$ not only characterizes a $d$d invertible 't Hooft anomaly of a symmetry $G$, 
but also characterizes a one-higher dimensional {\bf \emph{invertible topological field theory}} ({\bf \emph{iTFT}})
with a partition function $\bZ_{d+1}$ of the symmetry $G$. 
The $\bZ_{d+1}$ turns out to be a \emph{cobordism invariant} in $d+1$d 
(\cite{Witten2019bou1909.08775, Kapustin2014tfa1403.1467, Kapustin1406.7329, Wang1405.7689,
2016arXiv160406527F, WanWang2018bns1812.11967} and references therein). 
These iTFTs $\bZ_{d+1}$ form an abelian group: 
the group binary operation is the stacking of iTFTs, 
the identity element is 1, 
the inverse is $\bZ_{d+1}^{-1}$; 
they also obey closure, associativity, and commutativity. 
By \cite{2016arXiv160406527F}, this cobordism group is a specific version $\TP_{d+1}(G)$ for smooth manifolds with $G$-structure.

{\bf \emph{Deformation classes of QFTs}}:
We can use the symmetry to organize the quantum fields 
into the representations of the symmetry group,
and give powerful selection rules constraining the quantum dynamics.
Anomalies are invariant under continuous deformations of parameters of QFTs, 
including the renormalization group flow 
\cite{tHooft1979ratanomaly} (\cite{Cordova1905.09315, Cordova1905.13361} and references therein). 
Given the importance of 
symmetries and their associated 't Hooft anomalies,
Seiberg \cite{NSeiberg-Strings-2019-talk} and others  \cite{WangWen2018cai1809.11171}
conjectured that any QFTs with the same symmetry $G$ and same anomaly $\bZ_{d+1}$
can be deformed to each other by adding degrees of freedom at short distances
that preserve the same symmetry and maintain the same anomaly.
Namely, the whole system allows a large class of deformations via 
all symmetric interactions between the original QFT and other new higher-energy sectors of extra QFTs.
This organization principle connects a large class of QFTs together 
within the same data $(G, \bZ_{d+1})$
via any symmetric deformation (possibly with discontinuous or continuous quantum phase transitions \cite{subirsachdev2011book} between different phases).
The \emph{deformation class of QFTs} in $d$d \cite{NSeiberg-Strings-2019-talk},
is indeed related to the \emph{cobordism or deformation class of iTFTs} in one dimensional higher in $d+1$d \cite{2016arXiv160406527F}.

However, it is widely believed that there is no global symmetry in quantum gravity (QG) 
(\cite{HarlowOoguri1810.05338} and references therein).
Skeptical readers may hold prejudice against this 
deformation class labeled by $(G, \bZ_{d+1})$, possibly not universal nor useful in QG.
But several reasons below suggest its importance even in QG.
First, there could still be various QFTs with global symmetries localized on a brane or obtained via compactification from QG.
Second, McNamara-Vafa \cite{McNamara2019rupVafa1909.10355} suggests that
if a $d$d effective theory is obtained via the compactification from a quantum gravity in a total spacetime $D$d, 
and if the compact ${(D-d)}$d manifold is in a cobordism class $\bZ_{D-d}$, 
it either suggests the theory belongs to the Swampland in QG,
or suggests the $\bZ_{D-d}$ must be appropriately \emph{trivialized} to zero in the full suitably interpreted QG cobordism group $\Omega_{D-d}^{\rm QG}=0$
(e.g., the trivialization of cobordism class via introducing new defects to cancel each other's cobordism class).
This original $d$d theory can be connected by a finite-energy $(d-1)$d domain wall to nothing on the other side.
So we may say that \Refe{McNamara2019rupVafa1909.10355} justifies the use of {\bf\emph{deformation class of quantum gravity}}.
Third, although there is no global symmetry in QG, there can be either explicitly broken global symmetry or dynamically gauged symmetry 
(i.e., global symmetry is dynamically gauged as gauge symmetry, which we shall call Weyl's \emph{gauge principle} \cite{Weyl1929ZPhy}) in QG.
Therefore, for all the reasons above, it is promising to explore the idea of 
cobordism and deformation class further, in the context of both QFT \cite{NSeiberg-Strings-2019-talk} and QG \cite{McNamara2019rupVafa1909.10355}.

In this work, we shall focus on a 
specific question: {\bf\emph{What is the deformation class of 
the 4d Standard Model}} (SM) \cite{Glashow1961trPartialSymmetriesofWeakInteractions, Salam1964ryElectromagneticWeakInteractions, Salam1968, Weinberg1967tqSMAModelofLeptons}? 
Before attempting to solve this, we shall sharpen the question further.
We need to first specify the spacetime-internal symmetries of SM, typically written as
\bea
G\equiv ({\frac{{G_{\text{spacetime} }} \times  {{G}_{\text{internal}} } }{{N_{\text{shared}}}}}) \equiv {{G_{\text{spacetime} }} \times_{{N_{\text{shared}}}}  {{G}_{\text{internal}} } }.
\eea
The ${N_{\text{shared}}}$ is the shared common normal subgroup symmetry between ${G_{\text{spacetime} }}$ 
and ${{G}_{\text{internal}} }$, 
e.g. ${N_{\text{shared}}}$ can be the fermion parity symmetry $\Z_2^F$, which acts on fermions by $\psi \mapsto - \psi$.
The Lie algebra of the internal symmetry of SM is $su(3) \times su(2) \times u(1)$,
but the global structure Lie group $G_{\text{SM}_q}$ has four possible versions \cite{AharonyASY2013hdaSeiberg1305.0318, Tong2017oea1705.01853}:
\bea
G_{\SM_q} \equiv \frac{\SU(3) \times   \SU(2) \times \U(1)_{\tilde Y}}{\Z_q},  \quad \text{ with } q=1,2,3,6.
\eea
We also include the baryon minus lepton ${ \mathbf{B}-  \mathbf{L}}$ vector symmetry respected by the SM gauge structure and Yukawa-Higgs terms.
We consider both a continuous $\U(1)_{{ \mathbf{B}-  \mathbf{L}}}$ or $\U(1)_{X}$, or a discrete order-4 finite abelian $\Z_{4,X}$ group;
there $X \equiv 5({ \mathbf{B}-  \mathbf{L}})-\frac{2}{3} {\tilde Y}$ is Wilczek-Zee's chiral symmetry \cite{Wilczek1979hcZee} with the electroweak hypercharge ${\tilde Y}$.
%
We decide not to include extra discrete charge conjugation, parity, or time-reversal (C, P, or T) symmetries
as they are already individually broken at the SM energy scale.
The SM is a Lorentz invariant QFT that obeys the spacetime rotation and boost symmetry 
${G_{\text{spacetime} }} = \Spin$ group 
(Spin(3,1) in Lorentz signature and Spin(4) in Euclidean signature).
Gathering the above data, we learn that two physically pertinent versions of spacetime-internal symmetries of SM are:
\bea \label{eq:spacetime-internal}
G^{\U(1)}_{\SM_q} \equiv \Spin \times_{\Z_2^F} \U(1)_{} \times G_{\SM_q} \text{ and }
G^{\Z_{4,X}}_{\SM_q} \equiv  \Spin \times_{\Z_2^F} \Z_{4,X} \times G_{\SM_q}.
\eea
Both versions of the symmetry \eqq{eq:spacetime-internal} 
are compatible with the representation (rep) of the 3 families of 15 Weyl fermions of quarks and leptons 
(below written as left-handed 2-component Weyl spinors of the spacetime Spin group)
confirmed by experiments, 
possibly with or without the 16th Weyl fermion (the right-handed neutrino sterile to SM gauge force):
\bea \label{eq:SMrep}
&& \bar{d}_R \oplus {l}_L  \oplus q_L  \oplus \bar{u}_R \oplus   \bar{e}_R  
\oplus
n_{\nu_{j,R}} {\bar{\nu}_{j,R}}\cr
&&\sim 
(\overline{\bf 3},{\bf 1})_{2} \oplus ({\bf 1},{\bf 2})_{-3}  
\oplus
({\bf 3},{\bf 2})_{1} \oplus (\overline{\bf 3},{\bf 1})_{-4} \oplus ({\bf 1},{\bf 1})_{6}  \oplus n_{\nu_{j,R}} {({\bf 1},{\bf 1})_{0}}
\text{ for each family of } su(3) \times su(2) \times u(1)_{\tilde Y}.
\eea
We use $n_{\nu_{e,R}}, n_{\nu_{\mu,R}}, n_{\nu_{\tau,R}} \in \{ 0, 1\}$
to label either the absence or presence of the 16th Weyl fermion for each of the three families of electron $e$, muon $\mu$, or tauon $\tau$ types.
In general, we can consider $N_f$ families (typically $N_f=3$)
and a total sterile neutrino number $n_{\nu_{R}} \equiv \sum_j n_{\nu_{j,R}}$ 
{that can be equal, smaller, or larger than 3}.

In \Sec{sec:Ungauged}, 
we consider the 4d SM with internal symmetry $G_{\text{SM}_q}$ ungauged as a global symmetry group.
Both versions of spacetime-internal symmetries \eqq{eq:spacetime-internal}
are treated as global symmetries, so we study the global symmetries and their 't Hooft anomalies of the {\bf \emph{ungauged Standard Model}} at $d=4$.
However, 't Hooft anomaly of the ungauged theory
is the obstruction of gauging. 
Some \emph{'t Hooft anomalies of global symmetries}
directly descend to the \emph{dynamical gauge anomalies of gauge symmetries},
after gauging $G_{\text{SM}_q}$.
So we can also use the cobordism group $\TP_{5}(G)$ to check the \emph{anomaly cancellation consistency of the gauge theory}.

In \Sec{sec:gauged},
the internal symmetry $G_{\text{SM}_q}$ is thereafter dynamically gauged 
and treated as a gauge group in the 
{\bf \emph{gauged Standard Model}} with dynamical Yang-Mills gauge fields.  
We will see that the spacetime-internal symmetry \eq{eq:spacetime-internal} has to be modified to 
a new $G'$ 
by quotienting out the gauge group $[G_{\text{SM}_q}]$ and including extra generalized \emph{higher symmetry} \cite{Gaiotto2014kfa1412.5148}
(which is also an internal global symmetry but whose charged objects are extended 1d Wilson-'t Hooft line operators).
This $\TP_{5}(G')$ instead will be used to constrain the \emph{quantum dynamics of the gauge theory}.

\section{Standard Model with Internal Symmetry Ungauged and  
't Hooft Anomalies}
\label{sec:Ungauged}

We aim to determine the deformation class $(G, \bZ_{5})$
of ungauged 4d SM with spacetime-internal symmetries $G$ given in \eqq{eq:spacetime-internal}.
\Refe{Freed0607134, GarciaEtxebarriaMontero2018ajm1808.00009, WangWen2018cai1809.11171, 
WanWang2018bns1812.11967,
DavighiGripaiosLohitsiri2019rcd1910.11277, WW2019fxh1910.14668, HAHSIV} 
consider 
systematic anomaly classifications include 
several SM-like or Grand Unification (GUT)-like groups.
The relevant cobordism groups that classify
the 4d anomaly and 5d iTFT $\bZ_{5}$ are
\cite{WW2019fxh1910.14668, HAHSIV, JW2006.16996, JW2008.06499, JW2012.15860}
\bea
\TP_{5}(G^{\U(1)}_{\SM_q})= \TP_{5}(\Spin \times_{\Z_2^F} \U(1)_{} \times G_{\SM_q})
&=&
\TP_{5}(\Spin^c \times G_{\SM_q})
=
\Z^{11}.
\label{eq:TP5U1SM}
\\
\TP_{5}(G^{\Z_{4,X}}_{\SM_q}) 
=\TP_{5}(\Spin \times_{\Z_2^F} \Z_{4,X} \times G_{\SM_q})
&=&
\left\{ \begin{array}{ll} 
\Z^5\times\Z_2\times\Z_4^2\times\Z_{16}, & \quad q=1,3.\\
 \Z^5\times\Z_2^2\times\Z_4\times\Z_{16}, &\quad q=2,6.
 \end{array}
\right. 
\label{eq:TP5Z4SM}
\eea
Here are some remarks:

\noindent
\hspace{-4.2mm}
$\bullet$
We turn on
all possible \emph{background fields} (which are also called \emph{classical} \emph{non-dynamical} fields, or \emph{symmetry twists} \cite{Wang1405.7689},
that are \emph{not} summed over in the path integral)
that couple to the charged objects of QFT. 
Then we can perform all symmetry transformations of QFT by changing the values of background fields
to detect all invertible $\bZ_{5}=\e^{\ii \Theta}$.

\noindent
\hspace{-4.2mm}
$\bullet$
The symmetry $G$ consists of \emph{internal symmetry} and \emph{spacetime symmetry}.
{Internal symmetry} transformations act on charged objects that form a representation (rep) of group $G_{\text{internal}}$.
We couple the 0d charged particle to a 
1d background gauge field $A$.
{Spacetime symmetry} transformations act on the spacetime coordinates or metrics by spacetime diffeomorphisms.
The anomalies are detected by turning on 
(i) internal symmetry's background gauge fields,
(ii) spacetime symmetry's background fields via diffeomorphism or via placing the theory on $G$-structured manifolds, 
or (iii) mixed internal-spacetime background fields.   
These anomalies are 
respectively called (i) \emph{gauge anomaly}, (ii) \emph{gravitational anomaly}, or (iii) \emph{mixed gauge-gravity} or \emph{gauge-diffeomorphism anomaly}.
These adjectives, \emph{gauge} or \emph{gravity}, characterize the nature of the background fields of those anomalies.

\noindent
\hspace{-4.2mm}
$\bullet$
Cobordism group $ \TP_{d+1}(G)$
 {not only} contains all finite subgroup $\mathbb{Z}_{\rm n}$ (the torsion part) of the
standard bordism group $\Omega_{d+1}^{G}$, {but also} contains the integer $\mathbb{Z}$ classes 
descended from those $\mathbb{Z}$ (the free part) of the bordism group $\Omega_{d+2}^{G}$.

\noindent
\hspace{-4.2mm}
$\bullet$
The $d$d \emph{perturbative local anomalies} \cite{Adler1969gkABJ, Bell1969tsABJ, AlvarezGaume1983igWitten1984}
correspond to the torsion free $\Z$ classes of $\TP_{d+1}(G)$. 
They are detected by infinitesimal gauge/diffeomorphism transformations 
and captured by Feynman diagram calculations.
\\
The $d$d \emph{nonperturbative global anomalies} \cite{Witten1982fp, Witten1985xe, WangWenWitten2018qoy1810.00844}
correspond to the torsion $\mathbb{Z}_{\rm n}$ classes of $\TP_{d+1}(G)$.
They are detected via large gauge/diffeomorphism transformations that \emph{cannot} be continuously deformed from the identity.
\\
These adjectives, \emph{perturbative or not}, characterize the \emph{local} or \emph{global} nature of those anomalies.\\

Now we can obtain the deformation class $\bZ_{5}$ of the ungauged SM by checking which 't Hooft anomaly persists:
\begin{enumerate}[leftmargin=-0mm]

\item For SM with a $\U(1)_{{ \mathbf{B}-  \mathbf{L}}}$ vector symmetry, 
\eq{eq:TP5U1SM} 
shows $\Z^{11}$ classes of local anomalies captured by triangle Feynman diagrams.
Given the SM chiral fermion content \eqq{eq:SMrep},
all of local anomalies cancel to zero \cite{Weinberg1996Vol2} except two $\Z$ classes \cite{JW2006.16996}:
the cubic pure gauge $\U(1)_{{ \mathbf{B}-  \mathbf{L}}}^3$ and 
mixed gauge-gravity $\U(1)_{{ \mathbf{B}-  \mathbf{L}}}$-(gravity)$^2$ anomalies.
Their 5d iTFT evaluated on a 5d manifold ${M^5}$ is
\bea \label{eq:Z5U1BL}
\bZ_{5}^{\U(1)_{}} 
\equiv \exp(\ii 
(-N_f+n_{\nu_{R}} )
((\int_{M^5} {A} c_1^2) + \frac{1}{48}  \rCS_3^{T(\PD(c_1))})
),
\eea
depending on the family number $N_f$ and total sterile neutrino number $n_{\nu_{R}}$.
Typically for $N_f=3$, we have
$$-N_f+n_{\nu_{R}} \equiv 
-N_f+
\sum_j 
 n_{\nu_{j,R}}
= -3+n_{\nu_{e,R}} +  n_{\nu_{\mu,R}} + n_{\nu_{\tau,R}} + \dots.$$
Let us explain the components in \eqq{eq:Z5U1BL}.
The $A$ is the $\Spin \times_{\Z_2^F} \U(1)_{{ \mathbf{B}-  \mathbf{L}}} \equiv \Spin^c$ gauge field, while $2A$ is the ordinary abelian gauge field,
whose first Chern class $c_1=\frac{\dd (2 A)}{2 \pi}$ satisfies a spin-charge relation 
$c_1= w_2(TM) \rmod 2$
\cite{SeibergWitten1602.04251}
with the second Stiefel-Whitney class $w_2(TM)$ of spacetime tangent bundle $TM$ of the base manifold $M$.
The gravitational Chern-Simons 3-form is 
$\rCS_3^{TM^3} \equiv \frac{1}{4 \pi} \int_{M^3= \prt M^4} \Tr( \omega \dd \omega + \frac{2}{3} \omega^3)
=\frac{1}{4 \pi} \int_{M^4} \Tr( R(\omega) \wedge R(\omega))$ where $M^3$ is evaluated as a boundary $\prt M^4$,
{while $\omega$ is the 1-connection of tangent bundle $TM$ and $R(\omega)$ is the Riemann curvature 2-form of $\omega$}.
In \eq{eq:Z5U1BL}, we take ${M^3= \PD(c_1)}$ to be a 3-manifold Poincar\'e dual (PD) to the degree-2 first Chern class $c_1$ on the 5d $M^5$.
On a closed oriented $M^4$, we further have $\frac{1}{4 \pi} \int_{M^4} \Tr( R(\omega) \wedge R(\omega)) = {2 \pi} \int_{M^4} p_1 (TM) = {2 \pi} \cdot 3  \sigma$,
where $p_1$ is the first Pontryagin class of $TM$ and the $\sigma$ is the $M^4$'s signature. 
Overall, 
the 5d term \eqq{eq:Z5U1BL} descends from a 6d anomaly polynomial on $M^6$ at the U(1)-valued $\theta = 2 \pi$: 
$$
\exp(\ii \theta 
(-N_f+n_{\nu_{R}} )
( (\int_{M^6} \frac{1}{2}  c_1^3) + \frac{1}{16}  \sigma(\PD(c_1)))
).$$
Since $\U(1)_{Y}$ is fully anomaly-free, our above result is applicable to
SM with a $\U(1)_{X \equiv 5({ \mathbf{B}-  \mathbf{L}})-\frac{2}{3} {\tilde Y}}$ chiral symmetry \cite{Wilczek1979hcZee};
we still obtain the same deformation class \eqq{eq:Z5U1BL} 
as $\bZ_{5}^{\U(1)_{}}$ when replacing $\U(1)_{ \mathbf{B}-  \mathbf{L}}$ to $\U(1)_X$.

\item For SM with a $\Z_{4,X}$ chiral symmetry, 
given the 
fermion content \eqq{eq:SMrep},
all $\Z^5$ classes of local anomalies in \eqq{eq:TP5Z4SM}
cancel to zero. 
%
All extra $\Z_2$ and $\Z_4$ classes of global anomalies in \eqq{eq:TP5Z4SM} also cancel,
 but the $\Z_{16}$ mixed gauge-gravity global anomaly can be nonzero \cite{JW2006.16996, JW2012.15860},
 whose corresponding 5d iTFT
 is
\bea \label{eq:Z5Z4X}
\bZ_{5}^{\Z_{4,X}} 
\equiv \exp(\ii 
(-N_f+n_{\nu_{R}} )
\frac{2\pi \ii}{16}   \eta_{4{\dd}}(\text{PD}( A_{{\Z_{2,X}}} )) 
).
\eea
The background gauge field $A_{{\Z_{2,X}}}$ is the generator of the cohomology group $\H^1(\B{\Z_{2,X}}, \Z_2)$ with ${\Z_{2,X}} \equiv {\Z_{4,X}}/{\Z_2^F}$
of ${\Spin \times_{\Z_2^F} {\Z_{4,X}}}$-structure manifold $M$.
The classifying space $\B^{n}\rA  \equiv K(\rA,n)$ is the Eilenberg-MacLane space for an abelian group $\rA$.
Hereafter we use the standard convention, 
all cohomology classes are pulled back to the manifold $M$ along the maps given in the definition of cobordism groups,
thus $\H^1(\B{\Z_{2,X}}, \Z_2)$ can be pulled back to $\H^1(M, \Z_2)$.
Here a 5d $\Z_{16}$ class Atiyah-Patodi-Singer (APS \cite{Atiyah1975jfAPS}) eta invariant 
 $\eta_{5{\dd}} = \eta_{4{\dd}}(\text{PD}( A_{{\Z_{2,X}}}))$ 
is written as the 4d eta invariant ${\eta}_{4{\dd}} \in \Z_{16}$ 
on the Poincar\'e dual (PD) submanifold of $A_{{\Z_{2,X}}}$ (i.e., 4d ${\Pin^+}$-manifold),
based on a Smith homomorphism for the bordism group 
$\Omega_5^{\Spin \times_{\Z_2} \Z_4} \cong \Omega_4^{\Pin^+} =\Z_{16}$
\cite{2018arXiv180502772T, Hsieh2018ifc1808.02881, GuoJW1812.11959, Hason1910.14039}.
The $\Z_{16}$ class 4d eta invariant ${\eta}_{4{\dd}}$ is the topological invariant of the interacting fermionic topological superconductor
of condensed matter in three spatial dimensions 
with an anti-unitary time-reversal symmetry $\Z_4^{T}$ such that 
the time-reversal symmetry generator $T$ squares to the fermion parity operator, namely $T^2=(-1)^F$ (see a review \cite{Senthil1405.4015, 1711.11587GPW}).
In contrast, in SM, we have the unitary ${\mathbf{B}-  \mathbf{L}}$-like symmetry ${\Z_{4,X}}$ whose generator $X$ squares to $X^2=(-1)^F$.


\item 
Thanks to the Lie group embedding $(\Spin \times_{\Z_2^F} \Z_{4,X} \times G_{\SM_6}) \subset  \Spin \times_{\Z_2^F} \Spin(10) \equiv G^{\Spin(10)}$ \cite{WW2019fxh1910.14668},
we can embed the 16-Weyl-fermion SM$_6$ (i.e., 
$n_{\nu_{e,R}}, n_{\nu_{\mu,R}}, n_{\nu_{\tau,R}}=1$) 
with rep {\bf 16} into a Spin(10) chiral gauge theory, the $so(10)$ GUT. 
We can again ask the deformation class
$(G^{\Spin(10)}, \bZ_{5}^{\Spin(10)})$
of the ungauged 4d model.
Its cobordism group
\cite{WangWen2018cai1809.11171, 
WanWang2018bns1812.11967,
WW2019fxh1910.14668} 
is a ${\rm p} \in \Z_2$ class for $\Spin(n)$ with $n \geq 7$:
 \bea 
 \label{eq:TP5Spin}
\TP_5(\Spin \times_{\Z_2^F}  \Spin(n \geq 7)) &=& \Z_2, \\
\label{eq:Z5Spin}
\bZ_{5}^{\Spin(n)}
\equiv {\exp(\ii \pi \; {\rm p}  \int_{M^5} w_2w_3)}=
{\exp(\ii \pi \; {\rm p} \int_{M^5} w_2(TM)w_3(TM))} &=&\exp(\ii \pi \; {\rm p}  \int_{M^5} w_2(V_{\SO(n)})w_3(V_{\SO(n)}) ).
\eea
Here $w_j(TM)$ or $w_j(V_{\SO(n)})$ is the $j$th Stiefel-Whitney class of the
spacetime tangent bundle $TM$, or the associated vector bundle of the principal gauge bundle of ${\SO(n)}=\Spin(n)/\Z_2^F$.
The only 5d iTFT represents a ${\rm p} \in \Z_2$ class 4d mixed gauge-gravity global anomaly.
The conventional $so(10)$ GUT \cite{Fritzsch1974nnMinkowskiUnifiedInteractionsofLeptonsandHadrons} 
has no 4d anomaly thus resides in a ${\rm p}=0$  trivial deformation class $\bZ_{5}^{\Spin(10)}=1$ 
\cite{WangWen2018cai1809.11171, WangWenWitten2018qoy1810.00844, WanWangWen2112.12148}. 
Ref.~\cite{Wang2106.16248, WangYou2111.10369GEQC} construct a modified $so(10)$ GUT 
with an extra discrete torsion class of 4d-5d Wess-Zumino-Witten term from the GUT-Higgs field.
The modified $so(10)$ GUT has a $w_2w_3$ anomaly, thus it resides in a ${\rm p}=1$ deformation class.
Since $\Spin(18) \supset \Spin(10)$ contains the same $w_2w_3$ deformation class,
one can work out a similar story for a Spin(18) chiral gauge theory or $so(18)$ GUT \cite{WilczekZee1981iz1982Spinors}.
In fact, within either ${\rm p} \in \Z_2$ deformation class, Ref.~\cite{Wang2106.16248, WangYou2111.10369GEQC} show
that the SM can quantum-phase transition to the GUT-like models 
of $su(5)$ \cite{Georgi1974syUnityofAllElementaryParticleForces}, 
$su(4) \times su(2) \times su(2)$ \cite{Pati1974yyPatiSalamLeptonNumberastheFourthColor}, 
or flipped $u(5)$ \cite{Barr1982flippedSU5} models. Especially for ${\rm p} =1$ deformation class, 
Ref.~\cite{Wang2106.16248, WangYou2111.10369GEQC} suggest 
a \emph{gapless deconfined quantum critical region} (analogous to \cite{SenthildQCP0311326}) with beyond-SM phenomena near the SM vacuum
by tuning the GUT-Higgs potential.
\end{enumerate}
 
In summary, the deformation class $(G,\bZ_5)$ 
of ungauged SM with $\U(1)_{{ \mathbf{B}-  \mathbf{L}}}$ or $\U(1)_X$
is $(G^{\U(1)_{}}_{\SM_q},\bZ_{5}^{\U(1)_{}} )$,
and that of ungauged SM with $\bZ_{5}^{\Z_{4,X}}$ 
is $(G^{\Z_{4,X}}_{\SM_q},\bZ_{5}^{\Z_{4,X}} )$, for any $q=1,2,3,6$.
For the ungauged SM embeddable into a conventional or modified $so(n)$ GUT, 
we obtain an extra deformation class data $(G^{\Spin(n)} , \bZ_{5}^{\Spin(n)} )$.
%
%

 \section{Standard Model with Internal Symmetry Gauged and with Higher Symmetries}

\label{sec:gauged}

Next we determine the deformation class of $G_{\SM_q}$-dynamically gauged SM.
Before solving that problem, 
let us recall the relationship between the deformation classes of the {ungauged} SM (\Sec{sec:Ungauged}) and the gauged SM:

\noindent
\hspace{-4.5mm}
$\bullet$
The vanishing 't Hooft anomalies (thus not presented in the deformation class $\bZ_5$)
for the symmetry that we will gauge in the ungauged SM
 indeed one-to-one correspond to the  \emph{dynamical gauge-diffeomorphism anomaly cancellation} conditions
for the gauged SM. 

\noindent
\hspace{-4.4mm}
$\bullet$
If the gauged SM has only $G_{\SM_q}$ gauged, while keeping the global symmetry $\U(1)_{{ \mathbf{B}-  \mathbf{L}}}$, $\U(1)_X$, or ${\Z_{4,X}}$ ungauged,
then this SM deformation class still includes
\eq{eq:Z5U1BL}, \eqq{eq:Z5Z4X}, and \eqq{eq:Z5Spin}. 

\noindent
\hspace{-4.2mm}
$\bullet$ If the gauged SM has not only $G_{\SM_q}$ but also ${ \mathbf{B}-  \mathbf{L}}$ or $X$ dynamically gauged,
then the 5d iTFT also has to be dynamically gauged to a new \emph{noninvertible} theory, 
which is either a noninvertible gapped  TFT or a gapless theory. 
In this scenario, we could no longer study the 4d SM on its own. The 4d SM and 5d theories are strongly correlated and coupled with each other 
through ${ \mathbf{B}-  \mathbf{L}}$ or $X$ gauge forces mediating long-range entanglements.

\noindent
\hspace{-4.2mm}
$\bullet$
In a gauge theory,
the {internal symmetry} also includes \emph{higher $n$-symmetries} as generalized global symmetries \cite{Gaiotto2014kfa1412.5148}
that act on $n$d charged objects (1d lines, 2d surfaces, etc.)
by coupling the $n$d objects to $n+1$d background gauge fields.
We denote the higher $n$-symmetry group $G$ as $G_{[n]}$.
The $n+1$d gauge field lives in a higher classifying space $\B^{n} G$ (namely, the Eilenberg-MacLane space)
of the group $G_{[n]}$. %
Once $G_{\text{SM}_q}$ is gauged, 
\emph{kinematically}, we may obtain extra 1-form electric and magnetic symmetries, $G_{[1]}^e$ and $G_{[1]}^m$,
which are determined from the center of gauge group $Z(G_{\SM_q})$,
and the Pontryagin dual of the homotopy group $\pi_1(G_{\SM_q})^{\vee}$.
Moreover the 0d SM gauge charge fermionic particle \eqq{eq:SMrep} explicitly breaks the $G_{[1]}^e$ from $Z(G_{\SM_q})$ to a subgroup 
\cite{Wan2019sooWWZHAHSII1912.13504, AnberPoppitz2110.02981, WangYou2111.10369GEQC, MonteroWang}
that acts trivially on all particle representations:
\bea \label{eq:1-symmetry}
\begin{tabular}{|c | c |c| c| c |}
\hline
		 & $Z(G_{\SM_q})$ &  $\pi_1(G_{\SM_q})^{\vee}$ &
		1-form $e$ sym $G_{[1]}^e$  &
		1-form $m$ sym $G_{[1]}^m$ 
		\\ 
		\hline
		 $G_{\SM_q} \equiv \frac{{\SU(3)} \times {\SU(2)} \times \U(1)_{\tilde{Y}}}{\Z_q}$ & 
		 $\Z_{6/q} \times \U(1)$ & $\U(1)$ &    $\Z_{6/q,[1]}^e$ & ${\U(1)}_{[1]}^m$  \\
		\hline
\end{tabular}.
\eea
Thus, we have to modify \eqq{eq:spacetime-internal}, by removing $G_{\SM_q}$ and including $G_{[1]}^e \times G_{[1]}^m$, to
\bea \label{eq:spacetime-internal-gauged}
G'^{\U(1)_{}}_{\SM_q} \equiv \Spin \times_{\Z_2^F} \U(1) \times \Z_{6/q,[1]}^e \times \U(1)_{[1]}^m \text{ and }
G'^{\Z_{4,X}}_{\SM_q} \equiv  \Spin \times_{\Z_2^F} \Z_{4,X} \times \Z_{6/q,[1]}^e \times \U(1)_{[1]}^m.
\eea
We denote ${q'} \equiv 6/q$, 
where ${q}=1,2,3,6$ maps to ${q'}=6,3,2,1$, and
define ${q'} \equiv {2^{n_2} \cdot 3^{n_3}}$ where ${q'}=1,2,3,6$ corresponds to $(n_2,n_3)=(0,0),(1,0),(0,1),(1,1)$.
Below we examine 
what are additional new terms in the deformation class of SM
due to 't Hooft anomalies of higher symmetries of 1d line operators.
All possible 4d invertible anomalies are classified by the 5d cobordism groups $\TP_5$ of \eqq{eq:spacetime-internal-gauged} \cite{HAHSIV}:
\bea
&&
\label{eq:higher-SM-U1}
\TP_5(G'^{\U(1)_{}}_{\SM_q}) = {\TP_5(\Spin \times_{\Z_2^F} \U(1)_{} \times \B \Z_{q'} \times \B\U(1))=\TP_5(\Spin^c \times \B \Z_{q'} \times \B\U(1))=\Z^2 \times \Z_{q'} },\cr
&&\;\quad \text{generated by 
${A} c_1^2$, \; $\frac{1}{48}  \rCS_3^{T(\PD(c_1))}$, \;
and $B^e_{\Z_{q',[1]}}   (\dd B^m_{\U(1),[1]} \rmod {q'})$.} \quad\quad\\
&&
\label{eq:higher-SM-Z4}
\TP_5(G'^{\Z_{4,X}}_{\SM_q}) = \TP_5(\Spin \times_{\Z_2^F} \Z_{4,X} \times \B \Z_{q'} \times \B\U(1))
=\Z_{16} \times (\Z_{4})^{n_2} \times  \Z_{q'}
,\cr
&&\;\quad \text{generated by $\eta_{4{\dd}}(\PD({A_{\Z_{2,X}}}))$, \; $\eta'_{1{\dd}} (\PD(\cP(B^e_{\Z_{2,[1]}})))$, \;
and $B^e_{\Z_{q',[1]}}   (\dd B^m_{\U(1),[1]} \rmod {q'})$.} \quad\quad
\eea
Below we   
first explain the 5d iTFTs of \eqq{eq:higher-SM-U1} and \eqq{eq:higher-SM-Z4},
and then determine the deformation class of the gauged SM:
\begin{enumerate}[leftmargin=-0mm]
\item  The $\Z^2$ and $\Z_{16}$ classes give the same 5d iTFTs already appeared in \Sec{sec:Ungauged}.
So the deformation class of ungauged SM in \eq{eq:Z5U1BL} and \eqq{eq:Z5Z4X}
still remains in that of gauged SM.

\item 
$B^e_{\Z_{q',[1]}}$ and $B^m_{\U(1),[1]}$ are the background fields of 1-form electric and magnetic symmetries respectively, 
namely ${\Z_{q',[1]}^e}$ and  ${\U(1)}_{[1]}^m$ in \eqq{eq:1-symmetry}.
Again, all cohomology classes are pulled back to the manifold $M$.
Since the maps $M\to \B^2\Z_{q'}$ and $M\to \B^2\U(1)$ are given in the definitions of the cobordism groups, 
the cohomology classes $B^e_{\Z_{q',[1]}} \in \H^2(\B^2\Z_{q'},\Z_{q'})$ and $B^m_{\U(1),[1]}\in\H^2(\B^2\U(1),\U(1))$ are pulled back to $\H^2(M,\Z_{q'})$ and $\H^2(M,\U(1))$ respectively.

\item {$\eta'_{1{\dd}} (\PD(\cP(B^e_{\Z_{2,[1]}})))$:}
$\eta'_{1{\dd}}$ is a generator of the cobordism group $\TP_1(\Spin \times_{\Z_2^F} \Z_4^X)= \Z_4$.
{The} Pontryagin square 
 $\cP(B^e_{\Z_{2,[1]}}) \equiv B^e_{\Z_{2,[1]}} \smile B^e_{\Z_{2,[1]}} + B^e_{\Z_{2,[1]}}\smile_1 \delta B^e_{\Z_{2,[1]}}$
is a cohomology class in $\H^4(M,\Z_4)$, while $B^e_{\Z_{2,[1]}}$ is a cohomology class in $\H^2(M,\Z_2)$,
the $\smile_j$ is the $j$-th cup product {between cochains (we make the {$0$-th} cup product {$\smile$} implicit intentionally throughout the article)}, 
and $\delta$ is a coboundary operator. 
Since $M$ is $\Z$-orientable hence $\Z_4$-orientable, {the} Poincar\'e dual $\PD(\cP(B^e_{\Z_{2,[1]}}))$ is represented by a 1d $\Spin\times_{\Z_2}\Z_4$-submanifold of $M$, 
thus $\eta'_{1{\dd}}(\PD(\cP(B^e_{\Z_{2,[1]}})))$ is well-defined.
Whether the gauged SM has this mixed anomaly between the chiral $\Z_{4,X}$ and the ${\Z^e_{q',[1]}}$ symmetries 
(only at $q'=2,6$ thus $q=3,1$) can be computed 
via the techniques in \cite{Shimizu1706.06104, Tanizaki1711.10487, Cordova2018acb1806.09592DumitrescuClay, Wan2019oaxWWHAHSIII1912.13514}.

\Refe{Cordova2018acb1806.09592DumitrescuClay} finds that two adjoint Weyl fermions of SU(2) gauge theory with 
$\Spin \times_{\Z_2^F} \Z_{8} \times {\Z_{2,[1]}^e}$ symmetry has an anomaly index $1 \in \Z_4$ class of $A' \cP(B^e_{\Z_{2,[1]}})$ due to fractional SO(3) instantons, where
$A' \in \H^1(\B({\frac{\Z_{8}}{\Z_2^F}}), \Z_4)$ is pulled back to $\H^1(M,\Z_4)$.
In comparison, we consider that 
\ccblue{each family of SM has four SU(2) fundamental Weyl fermions as four SU(2) doublets {\bf 2}
gauged under $\SU(2) \times \U(1)$ gauge theory 
for $G_{\SM_q}$ at $q=1,3$}
with $\Spin \times_{\Z_2^F} \Z_{4,X} \times {\Z_{2,[1]}^e}$ symmetry. But this theory does not carry an anomaly index 
in the $\Z_4$ class of {$\eta'_{1{\dd}} (\PD(\cP(B^e_{\Z_{2,[1]}})))$},
because the $\U(1)_X$ symmetry is \emph{not} broken down to $\Z_{4,X}$ due to SU(2) instantons.
\ccblue{So any fractional $\U(2)$ instantons cannot induce any further mixed anomaly between the $\Z_{4,X}$-${\Z^e_{q',[1]}}$ symmetry.} 

%
\item $B^e_{\Z_{q',[1]}}   (\dd B^m_{\U(1),[1]} \rmod {q'})$:
This 5d term represents a mixed anomaly between the
electric 1-form $\Z_{q'}$ symmetry and the magnetic 1-form $\U(1)$ symmetry that we will detect in the 4d SM.
To define this 5d $B \dd B'$-like term mathematically precisely, 
we first define a generator $\tilde{\tau}_3  \equiv \dd B^m_{\U(1),[1]}  \in \H^3(\B^2\U(1),\Z)$ 
represented by the identity map $\B^2\U(1)=\B^3\Z\to \B^3\Z = \B^2\U(1)$. 
The $\tilde{\tau}_3$ is pulled back to $\H^3(M,\Z)$ along the map $M\to \B^2\U(1)$ given in the definition of the cobordism group.
{Here we regard $B^m_{\U(1),[1]}$ as the generator of $\H^2(\B^2\U(1),\U(1))= \Z$.
Also, we regard $\dd$ as the boundary map in the long exact sequence 
$\cdots\to\H^2(-,\U(1))\xrightarrow{\dd}\H^{3}(-,\Z)\to\cdots$ induced from the short exact sequence $1\to\Z\to\R\to\U(1)\to1$.} Note that $\dd$ is not the coboundary map $\delta$ which sends a cochain to a coboundary.
The factorization $\Z_{q'} = (\Z_{3})^{n_3} \times (\Z_{2})^{n_2}$
implies 
$B^e_{\Z_{q',[1]}}   (\dd B^m_{\U(1),[1]} \rmod {q'}) = -2 n_3 B^e_{\Z_{3,[1]}}   (\dd B^m_{\U(1),[1]}\rmod 3) + 3 n_2 B^e_{\Z_{2,[1]}}  (\dd B^m_{\U(1),[1]} \rmod 2)$.
 We check that, 
 for $G_{\SM_q}$ with Weyl fermion content in \eqq{eq:SMrep},
there is indeed a 4d mixed higher anomaly 
of 1-form symmetries between $\Z_{6/q,[1]}^e$ and $\U(1)_{[1]}^m$.
 %
For any family, the 
gauged SM contains the anomaly index ${\rm p}' =1 \in \Z_{q'}$ of the 5d iTFT. 
However, if the $\U(1)_{\tilde Y}$ magnetic monopole exists, 
kinematically it can break $\U(1)_{[1]}^m$ down to a discrete subgroup or to none (thus ${\rm p}' =0$) at the higher-energy monopole scale;
nonetheless $\U(1)_{[1]}^m$ can still re-emerge at the lower-energy SM scale. 
In short, the value ${\rm p}' \in \Z_{q'}$ depends on both the $\U(1)_{\tilde Y}$ magnetic monopole and the energy scale.    

\item For a discrete ${\mathbf{B}-  \mathbf{L}}$ vector or $X$ chiral symmetry, 
we so far focus on $(\Spin \times_{\Z_2^F} \Z_{4,X} \times \dots 
)$ for the 
SM.
But we could have other alternatives, e.g., 
$(\Spin \times_{\Z_2^F} \Z_{2 {\rm m}} \times \dots 
)$ for ${\rm m}=2,3,4,\dots$ 
for a larger discrete $\Z_{2 {\rm m},X}$ symmetry \cite{Hsieh2018ifc1808.02881, GuoJW1812.11959}.
We could also include the discrete ${\mathbf{B} + \mathbf{L}}$ vector symmetry:  
the $\U(1)_{\mathbf{B} + \mathbf{L}}$ is explicitly broken down, 
to $\Z_{2N_f, \mathbf{B} + \mathbf{L}}$ by the Adler-Bell-Jackiw anomaly \cite{Adler1969gkABJ, Bell1969tsABJ}
due to SU(2) instantons \cite{BelavinBPST1975},
or to $\Z_{2N_f \#, \mathbf{B} + \mathbf{L}}$ with an extra $q$-dependent factor denoted as $\#$ due to fractional instantons \cite{AnberPoppitz2110.02981}. 
But we have excluded ${\mathbf{B} + \mathbf{L}}$ in our work because it is not a good symmetry for the $su(5)$ or other GUTs. 
We leave other scenarios for future directions \cite{WangWanYou2204.08393}.

\item In summary, 
out of the multiple $\Z$ classes of local anomalies and $\Z_n$ classes of global anomalies 
(counted as a total finite group of order $2 \cdot 4^2 \cdot 16 \cdot 2 \cdot 4 \cdot 6 = 2^{13} \cdot 3=24576$ from 
\eqq{eq:TP5U1SM}, \eqq{eq:TP5Z4SM}, \eqq{eq:TP5Spin},
\eqq{eq:higher-SM-U1}, and \eqq{eq:higher-SM-Z4}),
we determine the deformation class of ungauged or gauged SM, labeled 
by the family number $N_f$, 
total sterile neutrino number $n_{\nu_{R}}$, 
and magnetic monopole datum ${\rm p}'$, the mod $q\equiv 6/{q'}$ of $G_{\SM_q}$,
written as a 5d iTFT:
 \be \label{eq:Z5-1symmetry}
\bZ_{5}^{} 
\equiv \exp(\ii 
(-N_f +n_{\nu_{R}}  )
(\dots)
+
\frac{2\pi \ii  {\rm p}'}{{q'}} 
 \int_{M^5} B^e_{\Z_{q',[1]}}   (\dd B^m_{\U(1),[1]} \rmod {q'}) 
).\quad
\quad
\ee
Here the $(\dots)$ are the $\Z^2$ or $\Z_{16}$ class iTFT already detected in 
\eq{eq:Z5U1BL} and \eqq{eq:Z5Z4X}. 
For the ungauged SM, 
if we embed the SM into the modified $so(10)$ GUT \cite{Wang2106.16248, WangYou2111.10369GEQC},
we could add \eq{eq:Z5Spin} into \eqq{eq:Z5-1symmetry}.
But again, once the ${ \mathbf{B}-  \mathbf{L}}$, $X$, or Spin(10) are dynamically gauged,
the 4d SM and the 5d becoming \emph{noninvertible} theories are correlated via mediating ${\mathbf{B}-  \mathbf{L}}$, $X$, or Spin(10) gauge forces.

The last term of \eqq{eq:Z5-1symmetry} with ${\rm p}'/q'$ dependence imposes a dynamical constraint on the 4d quantum gauge theory of the gauged SM:
the high energy theory or low energy fate of this gauged SM must match the same 't Hooft anomaly. 
For example, the $B^e \dd B^m$ term suggests that either 
${\Z_{q',[1]}^e}$ or  ${\U(1)}_{[1]}^m$ or both are spontaneously broken \cite{Gaiotto2014kfa1412.5148},
or there is no symmetric gapped trivial vacuum. 
The $\Z_{16}$ class 4d anomaly of \eqq{eq:Z5Z4X} and \eqq{eq:Z5-1symmetry} imposes another dynamical constraint in 4d:
there is no symmetric gapped phase for the odd classes in $\Z_{16}$,
but there exists symmetric gapped phases 
(namely gapped without $\Z_{4,X}$ chiral symmetry breaking)
with low energy TQFTs
for the even classes in $\Z_{16}$  \cite{Wang2017locWWW1705.06728, Hsieh2018ifc1808.02881, Cordova2019bsd1910.04962, Cordova1912.13069}.

\item Ultra Unification (UU) \cite{JW2006.16996, JW2008.06499, JW2012.15860}
replaces some of ``right-handed sterile neutrinos'' (sterile only to $G_{\SM_q}$ gauge force, but not sterile to $X$ or {certain} GUT forces) 
 by new exotic gapped/gapless sectors (e.g., topological or conformal field theory)
or by gravitational sectors with topological origins,
via nonperturbative global anomaly cancellation and cobordism constraints. Namely
$(-N_f+ n_{\nu_{R}}+ \dots) =0$
in $\Z_{16}$ for $\Spin \times_{\Z_2^F} \Z_{4,X}$ symmetry,
where $\dots$ is the anomaly index of the new gapped/gapless sectors from 4d or 5d. 
It can be generalized to $\Spin \times_{\Z_2^F} \Z_{2^m,X}$ symmetry with the 
$\Z_{2^{m+2}}$ global anomaly constraint for $m \geq 2$ \cite{Hsieh2018ifc1808.02881, GuoJW1812.11959}.
If we remove the new gapped/gapless sectors out of UU, the UU also resides in an SM deformation class.
There is a rich interplay between SM, GUT, and UU, where {\bf\emph{topological quantum phase transitions}} occur between the 15 vs 16 Weyl fermion models.
The new sectors  \cite{JW2006.16996, JW2008.06499, JW2012.15860, Wang2106.16248, WangYou2111.10369GEQC}
(gapped or gapless, including quantum phases or phase transitions, and quantum critical regions), 
neighbor to the SM vacuum in the quantum phase diagram landscape and in the deformation class, naturally exhibit beyond SM phenomena
and thus may provide Dark Matter candidates. 
Gapless quantum criticality also challenges the spacetime concept near the highly entangled critical regions, 
which may affect the energy density or pressure of spacetime, thus may relate to Dark Energy.
The deformation between neutrinos and exotic gapped/gapless sectors
may shed new light on the quantum version of the equivalence principle on 
the inertia mass (of the quantum gauge theory origin) and gravitational mass (of the gravitational origin).

\end{enumerate}

\noindent
\emph{Acknowledgments} ---
JW thanks Mohamed Anber, Yuta Hamada, Miguel Montero, Cumrun Vafa, 
and Edward Witten for conversations.
We thank Alek Bedroya, Cameron Krulewski, and Leon Liu for feedback on the manuscript. 
JW is supported by 
Harvard University CMSA.
ZW is supported by the Shuimu Tsinghua Scholar Program.
YZY is supported by a startup fund at UCSD.\\




\onecolumngrid

\bibliography{BSM-GEQC.bib}

\end{document}